# Explore of exfoliable multifunctional high-*k* two-dimensional oxides

Yue Hu[1], Jingwen Jiang[1], Peng Zhang[1], Fuxin Guan[2], Da Li[1], Zhengfang Qian[1], Pu Huang[1] and Xiuwen Zhang[1,3,*]

[1] Key Laboratory of Optoelectronic Devices and Systems of Ministry of Education and Guangdong Province, College of Physics and Optoelectronic Engineering, Shenzhen University, 518060 Shenzhen, China.

[2] Department of Physics, University of Hong Kong, Hong Kong, China.

[3] xiuwenzhang@szu.edu.cn

*Current address: Renewable and Sustainable Energy Institute, University of Colorado, Boulder, Colorado 80309, USA.

As the continuing down-scaling of field-effect transistors (FETs) in more-than-Moore integrated circuits, finding new functional two-dimensional (2D) materials with a higher dielectric constant (high-*k*) serve as gate dielectrics is critical. Here, we identify dozens of binary 2D oxides by screening potentially exfoliable bulk metal oxides despite of their non-layered structures followed by simulation of the exfoliation process. For dynamically stable materials, we fully characterize their static dielectric constants and electronic structures, among which $GeO_2$(011)/(101)/(1-11) 2D oxides exhibit unusually high *k* values (85-99), being much higher than the *k* of the currently highly regarded 2D dielectrics $CaF_2$ (k ~6) and β-$Bi_2SeO_5$ (k ~22), together with band gap of 3.3 eV. We further design 2D high-*k* oxides/2D semiconductors (such as $MoS_2$) heterostructures, and determine by DFT calculations whether they can form Van der Waals interfaces to evaluate their compatibility as gate dielectrics in 2D FETs. In addition to dielectric properties, we also explore magnetic and mechanical properties of potentially exfoliable 2D oxides, revealing a number of functional materials that can be studied experimentally, notably including ferromagnetic half semiconductors, non-magnetic spintronic materials, flexible high-*k* 2D oxides, and auxetic monolayers.

Two dimensional (2D) materials were used as semiconducting channels or metallic electrodes in the 2D field-effect transistors (FETs) that have been the focus of materials science and device physics in the last decades[1-8]. The gate dielectrics of the 2D FETS have been realized by transferring thin films of bulk oxides onto 2D semiconductors[9], deposition of high-*k* oxides on 2D materials[10], or oxidation of the 2D semiconductors into their native oxides[11]. On the other hand, 2D high-*k* oxides could be easily transferred onto the 2D semiconducting channels, offering the opportunity to realize the pure-2D FET with its gate, source and drain electrodes, semiconducting channel, as well as gate dielectrics formed by 2D crystals. Furthermore, the potentially fruitful physical properties in the 2D gate dielectrics analogous to the mechanical metamaterials[12] or valleytronic systems[13] could significantly enrich the functionalities of the pure-2D FETs. However, the material space of currently known 2D oxides are rather limited[14-17].

An important progress towards searching for 2D materials including oxides were taken by Mounet et al. where 5,619 layered bulk precursors were screened from 108,423 experimentally known 3D compounds, and 1825 of these were identified to be potentially exfoliable using density functional theory (DFT) calculations, including water soluble hydroxides, rare-earth oxyhalides and transition metal oxides[18]. Obviously, layered bulk precursors are extremely limited in bulk material space. This severely limits data mining for 2D materials. Non-layered bulk precursors, by contrast, are abundant and untapped, so if they could be exfoliable, non-layered bulk precursors would become a vast and diverse source of 2D materials. Certainly, many efforts have been made to find the possible exfoliation method for non-layered structures, such as liquid metal-assisted exfoliation[17], cryogenic exfoliation[19] and gel-blowing exfoliation[20]. Even so, the library of non-layered 2D materials is still scarce and limited in size.

In this study, we rethink the precondition for physical exfoliation, i.e., whether there is such a crystal plane in non-layered bulk precursors, not only is its interlayer bonding relatively weak, but the difference between the interlayer bonding and the in-plane bonding also is large enough to exfoliate the plane from its bulk precursor, a case that would be dismissed as being totally uninteresting from the point of strong chemical bonding. To this goal, the first step, we develop a general geometric algorithm to identify the crystal planes that meet the above two criteria from non-layered 3D structures; the second step, we perform DFT calculations of the interplanar binding energies ($E_b$) for the planes obtained by the geometric consideration, and quantitatively identify their potential for exfoliation. This results in a combination of 73 2D oxides that could be easily or potentially exfoliated from their non-layered 3D precursors. Among them, 61 dynamically stable 2D oxides are further characterized in electronic, magnetic and mechanical properties, disclosing dozens of functional materials including flexible high-*k* 2D oxides, magnetic and non-magnetic spintronic semiconductors and auxetic monolayers.

Our selection procedure starts from 1,921 binary metal oxides in Materials Project[21]. We consider the most stable non-layered compounds known experimentally, and select only materials with no more than 10 atoms in the unit cell for computational efficiency. Besides, we remove the materials containing f-valence electrons (such as $CeO_2$, EuO, etc.) to avoid the possible computational difficulty by DFT. After these criteria, we further enlarge the list by making sure that the famous dielectric oxides $TiO_2$, $ZrO_2$ and $HfO_2$ are in the list. Finally, we arrive at a mixed set of 48 non-layered compounds, including a sizable portion of the transition metal, post–transition metal, alkali-metal and alkaline-earth metal oxides. For this set of 3D precursors, we perform a magnetic filtering and screen out 9 magnetic oxides, see Supplementary Table 1. At this point, we have done with the preparatory work. The relevant workflow diagram is summarized in Supplementary Figure 1.

We next illustrate the procedure for calculating $E_b$ in greater detail. Firstly, we apply the geometric criteria to identify promising crystal planes for exfoliation. The criteria are based on the packing degree in-plane and out-of-plane, identifying those crystal planes who both possess large interplanar spacing ($d_{hkl}$) and close packing in-plane, i.e., a relatively weak interplanar interaction, together with a marked difference between in-plane and interplanar bonding. As a rule-of-thumb, a close-packed plane often has a large interplanar spacing and the lowest surface energy, tending to be exposed in the surface during the crystal growth[22]. Then, we rotate the close-packed plane to the (001) plane and increase the lattice constant c in the Z direction step by step, during which the bulk geometry and all the atoms in the system are fully optimized. We schematically display the atomic structural evolution in the stretching process in Fig. 1(b, c): with stretching, the interplanar bond strength gradually weakens until they separate, followed by the surface reconstruction of each separated part, eventually forming a structure that does not change with stretching.

We find that in the stretching process from 3D precursor to separation, the system energy increases gradually for all materials considered, while in the surface reconstruction phase, there are two types of energy evolution. Specifically, if there is a significant structural deformation in the surface reconstruction process, such as the formation of new chemical bonds in-plane, then the system energy usually decreases significantly, as shown in Fig. 1(d). While if the 2D structure after the surface reconstruction can still be roughly found the counterpart from its 3D precursor, then the system energy in the reconstruction phase usually continue to increase and eventually converges, see Fig. 1(a), which is similar to the layered structures[23]. Furthermore, we consider the vdW forces in the stretching process by using DFT-D3 correction method of Grimme et al[24]. In order to get an accurate estimate of $E_b$, an assessment of the quality of DFT-D3 is necessary. In Supplemetary Fig. 2, we calculate $E_b$ for well-known layered materials including graphite, hexagonal boron nitride and several transition-metal dichalcogenides. The results show that the values obtained with the DFT-D3 functional are in good agreement with the reference calculations performed by DF2-C09 functional[18] and random phase approximation (RPA)[23].

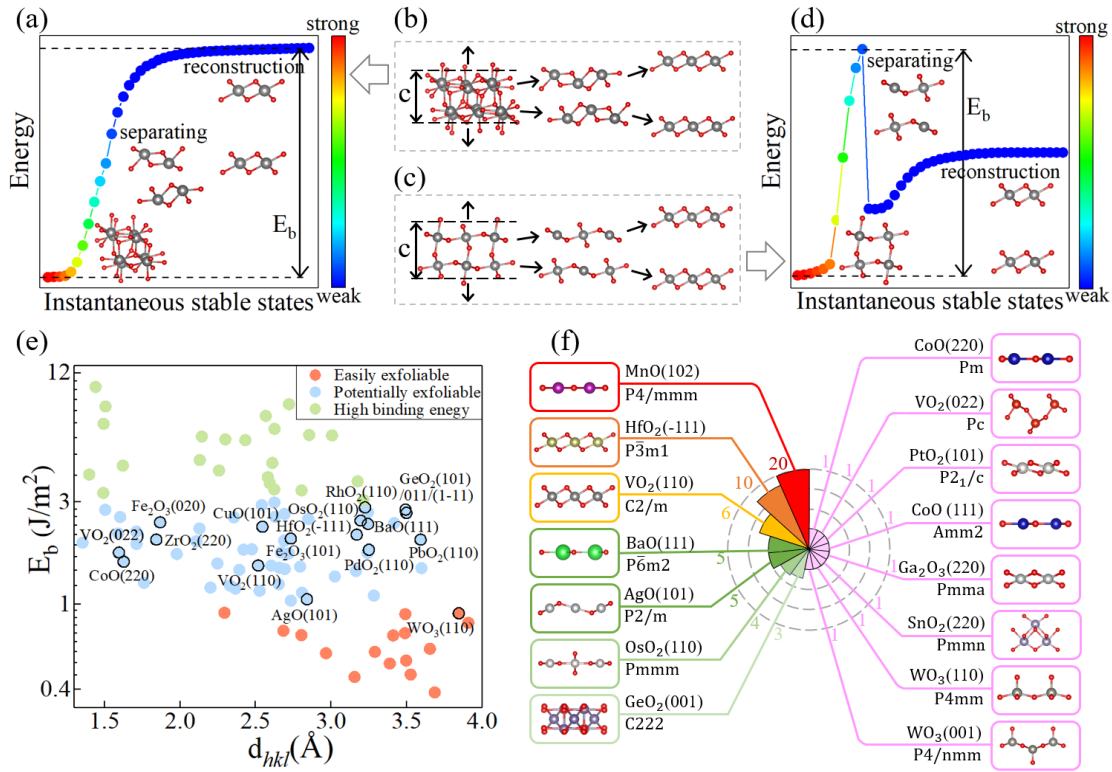

Fig.1 (a, d) Schematic illustration of interplanar binding energy ($E_b$) curves. (b, c) Procedure for calculating $E_b$ by increasing lattice constant c. (e) $E_b$ versus interplanar spacing ($d_{hkl}$). Materials classified as easily exfoliable, potentially exfoliable and high binding energy are demonstrated in different colors. Functional 2D oxides including high-*k*, spintronic and auxetic oxides are marked by chemical formulas. (f) Polar histogram for dynamically stable easily/potentially exfoliable 2D oxides classified by space group. The crystal structures and chemical formulae of representative materials for each space group are shown, as well as the total number of 2D oxides contained in each space group. Red balls denote O atoms and the other colored balls indicate metal atoms, which are all the same in the following figures.

In Supplementary Fig. 3, we compare $E_b$ of close-packed and nearly close-packed planes. The conclusion is that in general, close-packed planes have a lower $E_b$ than nearly close-packed planes, which is in excellent agreement with our expectation. Besides, there are several cases where nearly close-packed planes have a lower $E_b$. We find these cases have one thing in common. That is, they have a larger $d_{hkl}$ compared to that of close-packed planes, see Supplementary Table 2. In conclusion, we observe that when the crystal plane satisfies in-plane close-packing and large interplanar spacing simultaneously, it exhibits a low $E_b$, while if the crystal plane satisfies only one of the two conditions, further calculations to evaluate its potential for exfoliation is needed. Even so, using the close-packed degree in-plane combined with the spacing between adjacent planes to roughly predict the promising planes for exfoliation is still a fast and effective method.

We plot in Fig. 1(e) $E_b$ and $d_{hkl}$ for 100 2D oxides originated from 48 non-layered bulk precursors. The corresponding values are listed in Supplementary Table 2. Notably, the inverse relation is clearly noticeable when $E_b$ and $d_{hlk}$ are plotted. That is, most materials with low $E_b$ exhibit large $d_{hlk}$, distributing in the bottom right corner in Fig. 1(e); while most materials with high $E_b$ exhibit small $d_{hlk}$, concentrating in the top left corner in Fig. 1(e). This is expected, since large interplanar spacing means weak interplanar bonding. An extreme example, layered materials have relative large interplanar spacing, whose interlayer interactions are weak vdW forces accordingly.

To further identify exfoliable oxides, we choose thresholds of 1.0 J/m$^2$ for $E_b$, and classify materials falling below this threshold as "easily exfoliable", marked by red in Fig. 1(e). This choice identifies layered 2D materials commonly exfoliated in the experiments[25]. The results show that most binding energies are higher than this value for all materials under consideration. Even so, there are still 16 oxides belong to this region, among which we find one new and compelling 2D oxide WO$_3$(110), it has extraordinary mechanical property, i.e., the monolayer expands laterally, regardless if a longitudinal stretched or compressed is applied. In the top left corner of Fig. 1(e), a number of materials exhibit high $E_b$ and small $d_{hlk}$. This group, shown in green, can be clearly separated from the other groups, the boundary being set at 3 J/m$^2$. Above this value, materials are considered to be difficult to exfoliate, and are excluded from further investigation in physical properties. Between these two regions, the remaining materials (shown in blue) exhibit relatively weak bonding. We classify materials (57) belonging to this group as "potentially exfoliable" (in blue). 2D oxides HfO$_2$(-111), belonging to this region, has been synthesized in the experiments[17]. Besides, we find multiple 2D functional oxides in this region that have not been reported before, such as flexible high-$k$ oxides GeO$_2$(011)/(101)/(1-11), ferromagnetic half semiconductors VO$_2$(022), and so on.

Following, we assess the structural stability of 73 easily/potentially exfoliable 2D oxides by computing the phonon dispersion. We randomly select at least one structure for each chemical general formula from each type of space group, and calculate their phonon dispersions. The computed results shown in Supplementary Figs. 4-29 reveal that most phonon dispersions do not have imaginary frequencies, indicating that most 2D oxides are dynamically stable. In Fig. 1(f), we display dynamically stable 2D oxides (61) according to space group, and the details are summarized in Supplementary Tables 3-18. The most common space group is P4/mmm, containing MnO(102) and 21 similar structures, see Supplementary Table 15. Some space groups like P-3m1, P-6m2 and Pmmn can also be commonly found in the layered 2D materials[18,26]. Besides, the graphene-like monolayers[27-30], such as ZnO(002), BeO(002)/(110) and BaO(111), have already been suggested as exfoliable, supporting the robustness of the exfoliation procedure. Encouraged by the results, we fully characterize the electronic ground states of 61 dynamically stable 2D oxides, revealing most of 2D oxides (49) are semiconducting, including 9 magnetic oxides, and the rest (12) are metallic, see Supplementary Table 19.

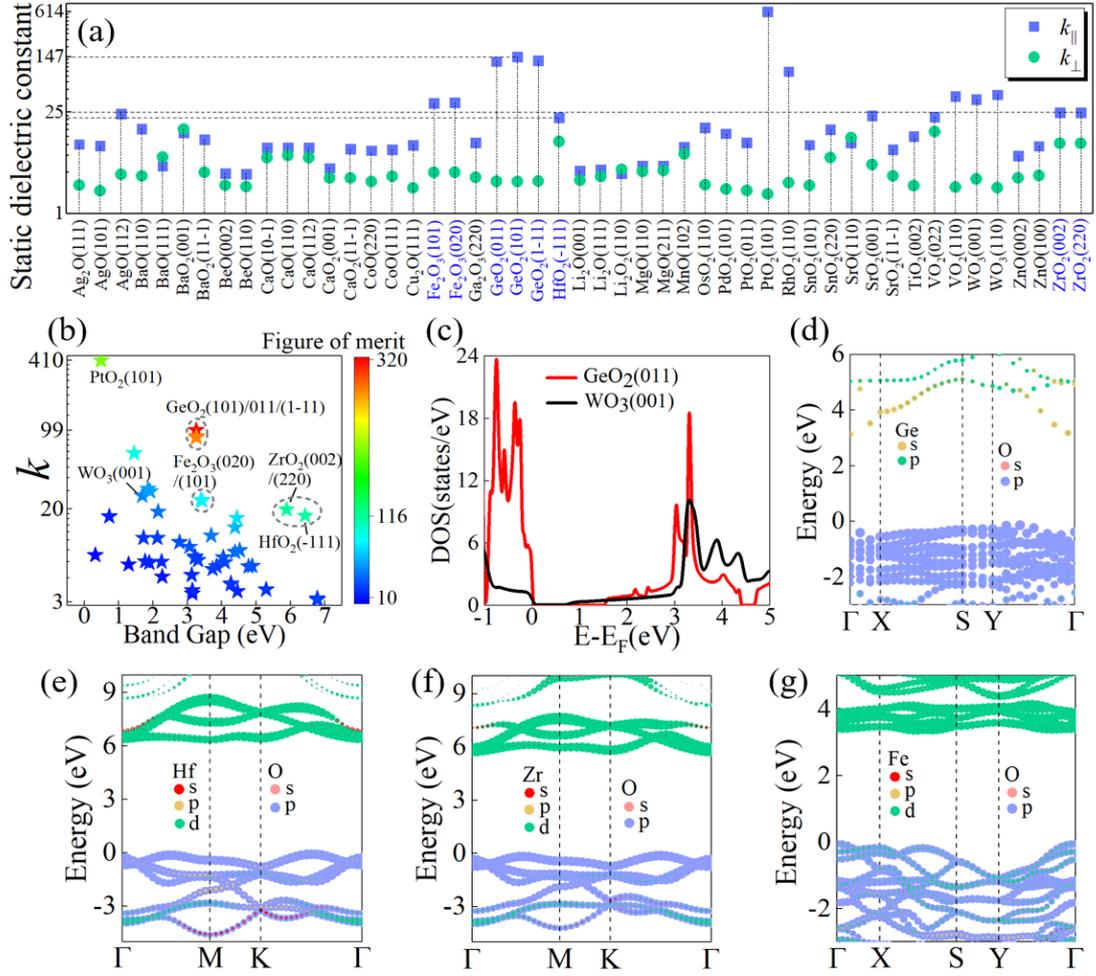

Fig. 2 Static dielectric properties for 2D semiconductor oxides: (a) In-plane ( ∥ ) and out-of-plane ( ⊥ ) static dielectric constants ($k$); (b) $k$ versus band gap ($E_g$, HSE), where each point is color coded according to the figure of merit ($E_g \cdot k$). Comparison chart for total DOS calculated at PBE level for 2D high-$k$ oxides (c) $GeO_2$(011) and $WO_3$(001). Partial band structures calculated with HSE functional for (d) $GeO_2$(011), (e) $HfO_2$(-111), (f) $ZrO_2$(002), (g) $Fe_2O_3$(020) monolayers.

To find new high-$k$ 2D materials, we calculate the static dielectric constants and band gaps of 2D semiconductor oxides (49). Fig. 2(a) displays the in-plane and out-of-plane static dielectric constants. Both the ionic and electronic contributions to the dielectric response are computed. The corresponding values are listed in Supplementary Table 20. The results show that $PtO_2$(101) and $GeO_2$(011)/(1-11)/(101) monolayers exhibit the highest in-plane dielectric constants (614, 127-147), which are higher than the highest in-plane dielectric constant of 2D TlF (98.4)[26] as far as we are aware; for the out-of-plane direction, the highest dielectric constant (15) belongs to $BaO_2$(001) monolayer. We plot the comparison chart for the ionic and electronic contribution both in the in-plane and out-of-plane directions in Supplementary Figs. 30-31. We observe that for most 2D oxides under consideration the in-plane dielectric response is dominated by the ionic contribution, while in the out-of-plane direction, the electronic component is higher than the corresponding ionic component. Our calculations also reveal that the ionic contribution differs greatly in-plane and out-of-plane, but not greatly for the electronic contribution (see supplementary Table 20), suggesting that the marked difference between the in-plane and out-of-plane static dielectric constants is mainly origin from the ionic contribution. The same is true for rare-earth oxyhalides[26].

We re-calculate band gaps ($E_g$) using HSE functional to get the accurate estimates of $E_g$, and plot the property map of $E_g$ versus dielectric constant ($k$) in Fig. 2(b). The $k$ is obtained by averaging the diagonal components of dielectric tensor. First, the inverse relation between $E_g$ and $k$ is roughly valid. This inverse relationship should be expected since the electronic susceptibility is inversely proportional to the energy difference of the transition states if first-order perturbation theory is taken into account, and the latter increases on average with the band gap. Secondly, the oxides possessing both large $E_g$ and $k$ are scarce. To select potentially high-$k$ 2D oxides, we define a figure of merit for reducing the leakage current, and discussed in the methods section.

Among all the 2D oxides in Fig. 2(b), the figure of merits of $GeO_2$(011)/(1-11)/(101) monolayers stand out particularly because of unusually high $k$ (85-99), being much higher than the $k$ of the currently highly regarded 2D dielectrics $CaF_2$ ($k$ ~6)[31] and β-$Bi_2SeO_5$ ($k$ ~22)[32], coupled with $E_g$ > 3eV, with possible applications in DRAM according to the international technology roadmap of semiconductors. The origin of the high-$k$ of $GeO_2$(011)/(1-11)/(101) is that a large number of states near or at the top/bottom of valence/conduction bands have a relatively large number of low-energy transition states. Therefore $GeO_2$(011)/(1-11)/(101) have high dielectric constants, even though their band gaps are relatively large, for example, compared to the $WO_3$(001) monolayer, see the total DOS comparison chart in Fig. 2(c). The orbital projected band structure of $GeO_2$(011) monolayer in Fig. 2(d) reveal that the states in the top of valence bands are dominated by O p orbitals, while Ge s and p orbitals contribute mainly to the states in the bottom of conduction bands.

In addition to $GeO_2$ (011)/(1-11)/(101), we also have identified several noteworthy candidates with figure-of merits above 80 and band gaps above 3 eV, including $ZrO_2$(002)/(220), $HfO_2$(-111) and $Fe_2O_3$(020)/(101) monolayers. The orbital projected band structures of $HfO_2$(-111), $ZrO_2$(002), and $Fe_2O_3$(020) monolayers computed at the HSE level shown in Fig. 2(e-g) reveal that the states near the bottom of conduction bands are dominated by the d orbitals of transition metals (Fe, Zr, Hf), while the main contribution to the states near the top of the valence bands mainly come from O p orbitals. Besides, $HfO_2$ thin film prepared using atomic layer deposition or other chemical and physical vapor-phase deposition techniques have been commonly used as gate dielectrics in 2D transistors according to the reports[33].

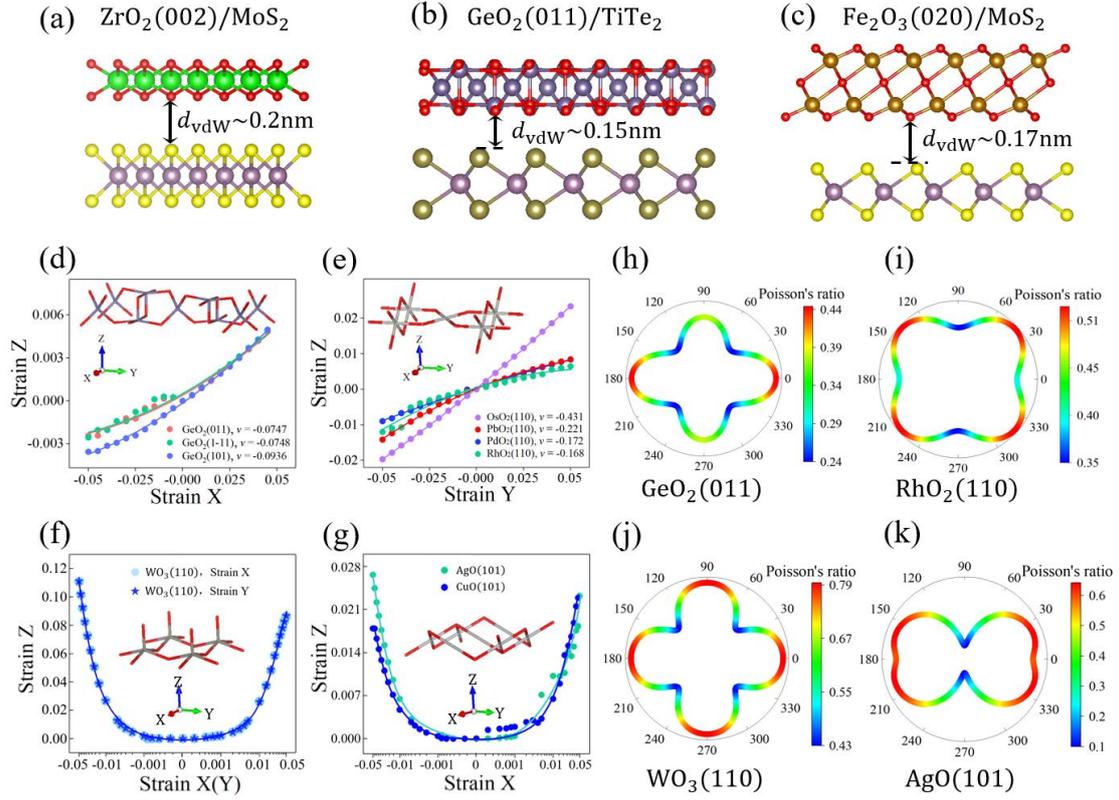

Fig. 3 Schematic illustrations for 2D high-$k$ oxides (up layer) and 2D semiconductors (down layer) integration: a clean vdW interface exists in between (a) $ZrO_2$(002) and $MoS_2$ (b) $GeO_2$(011) and $TiTe_2$, and (c) $Fe_2O_3$(020) and $MoS_2$, where the vdW gap ($d_{vdW}$) is indicated by arrows. Out-of plane mechanical response: strain in the Z direction induced by the uniaxial strain in the (d) X direction for 2D high-$k$ oxides $GeO_2$(011)/(1-11)/(101), (e) in the Y direction for $PdO_2$(110), $PbO_2$(110), $RhO_2$(110) and $OsO_2$(110) monolayers, (f) in the X/Y directions for $WO_3$(110) monolayer, and (g) in the X direction for AgO(011) and CuO(101) monolayers. Four insets show the crystal structures. (h-k) In-plane Poisson's ratio as a function of in-plane angle. 0 degree corresponds to the X axis.

Van der Waals integration, a bond-free integration strategy without lattice and processing limitations, has received wide spread attention. To judge whether the 2D high-$k$ oxides we obtained can form vdW interface with the 2D semiconductors ($MoS_2$ and $TiTe_2$), we model the "2D high-$k$ oxides/2D semiconductors" heterostructures, and display three examples in Fig. 3(a-c). We then fully optimized these structures. Our calculations show that the calculated vdW gaps ($d_{vdW}$) for different heterostructures exhibit similar values of ~0.2 nm (see Supplementary Table 21), comparable to that of the artificially assembled vdW interfaces, e.g., Au/$MoS_2$[34], BN/graphene[35], and $WSe_2$/$Bi_2Se_3$[36], demonstrating that these 2D high-$k$ oxides can be integrated with 2D semiconductors by Van der Waals interactions. By the way, we consider the heterostructure of $GeO_2$(011) and $TiTe_2$, due to the calculation burden caused by the oversized system of $GeO_2$(011)/$MoS_2$ heterostructure that contains more than two hundreds atoms.

Another advantage of 2D materials in the actual use, such as for the thin-film FET in the integrated circuits or for the optoelectronic applications, benefits from its unconventional mechanical properties (unique flexibility, super toughness, higher indentation resistance, etc.). We hence further investigate the mechanical properties of 2D oxides. Excitingly, we find 2D high-*k* oxides $GeO_2$(011)/(1-11)/(101) exhibit negative Poisson's ratio (NPR) behavior in the out-of-plane direction. The strain curves shown in Fig. 3(d) demonstrate that the material expands/contracts in the Z direction, if a stretched/compressed in the X direction is applied, with NPR values approximately three times higher than that of black phosphorus[12], indicating that high-*k* 2D oxides $GeO_2$(011)/(1-11)/(101) possess superior toughness and higher shear resistant.

Furthermore, we observe more obvious NPR behavior in the 2D oxides $PdO_2$(110), $PbO_2$(110), $RhO_2$(110), $IrO_2$(110) and $OsO_2$(110), whose NPR values are approximately 6~16 times that of black phosphorus[12], see Fig. 3(e). More interestingly, we observe a more peculiar mechanical response in 2D oxides $WO_3$(110), AgO(101) and CuO(101). Specifically, the monolayers expand in the Z direction, regardless if a stretched or compressed in the X/Y direction is applied, see Fig. 3(f, g). Up to now, we only find one report that Ma et al. discover similar behavior in the Pd-decorated borophene[37]. In contrast to this report, this behavior is intrinsic for $WO_3$(110), AgO(101) and CuO(101) monolayers, and originates from their unique puckered structures, see the insets in Fig. 3(f, g).

We offer a mechanistic explanation to understand the above special mechanical responses, which is summarized in the Supplementary Figs. 33-36. Furthermore, our calculated in-plane Poisson's ratio at arbitrary angle for these auxetic 2D oxides shown in Fig. 3(h-k) reveal that all these 2D oxides exhibit high anisotropy and positive Poisson's ratios in the in-plane direction.

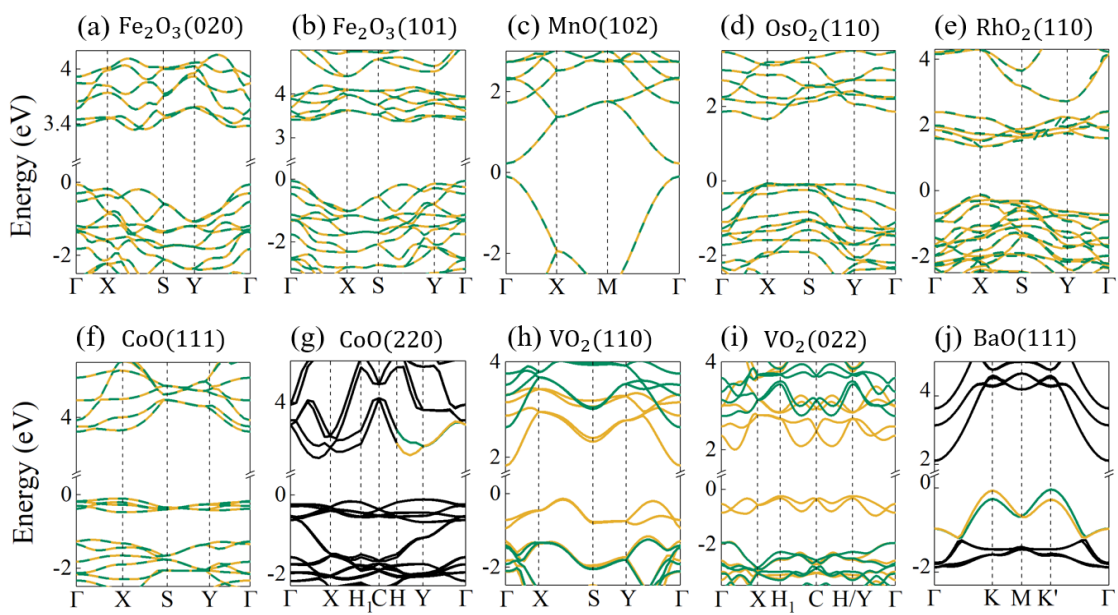

Fig. 4 Spin-polarized band structures calculated at HSE level for (a-g) antiferromagnetic (AFM) 2D oxides: $Fe_2O_3$(020), $Fe_2O_3$(101), MnO(102), OsO2(110), RhO2(110), CoO(111) and CoO(220), and for (h, i) ferromagnetic 2D oxides: $VO_2$(110) and $VO_2$(022). Band

structures with spin-orbital coupling calculated at HSE level for non-centrosymmetric 2D oxides (g) CoO(220) and (j) BaO(111). The regions near conduction band minimum of CoO(220) and valence band maximum of BaO(111) show a clear spin-splitting. Orange and green lines denote spin-up and spin-down polarization, respectively.

We next explain the magnetic and non-magnetic spintronic 2D oxides in detail. We fully characterized the magnetic ground-states of 11 easily/potentially exfoliable 2D oxides with stable phonon dispersions originated from 9 magnetic bulk precursors by comparing the energy of ferromagnetic (FM) and several antiferromagnetic (AFM) magnetic configurations. A total of 9 2D oxides are found to support non-trivial magnetic ordering, including 7 AFM and 2 FM oxides, and their magnetic configurations are shown in Supplementary Fig. 37. The spin-polarized band structures shown in Fig. 4(a-i) reveal that 9 magnetic oxides are semiconducting. Among them, 2D AFM oxides $Fe_2O_3$(020) and $Fe_2O_3$(101) exhibit high static dielectric constants and possess band gap of 3.4 eV. Multifunctional high-$k$ 2D oxides may be important in future, openning up new opportunities for fundamental research and achieving unprecedented device performance.

Furthermore, there are 2 FM half semiconductor oxides $VO_2$(110) and $VO_2$(022) with possible applications in spintronics and data storage[38]. Their valence bands and conduction bands are spin splitting with the valence band maximum (VBM) and conduction band minimum (CBM) possessing the same spin channel, as shown in Fig. 3(h, i). According to the atomic formal charge, each O atom should gain two electrons from the V atom ($[Ar]3d^34s^2$), so the one remaining valence electron on the V atom, resulting in a net spin moment of 1 μB per V atom. This number, together with lowest-energy magnetic configurations and magnetic moments for 9 magnetic 2D oxides are summarized in Supplementary Table 22. Besides, The partial DOS of $VO_2$(110) and $VO_2$(022) monolayers in Supplementary Fig. 38 reveals that their VBM and CBM are dominated by the spin-up states V d orbitals.

We know that spin-orbit coupling (soc) can induce spin splitting and spin polarization if the material has sufficiently low crystal symmetry, even in nonmagnetic materials[39]. We hence further characterize the band structures with soc for all non-centrosymmetric 2D oxides (marked in blue in Table 19). We screen out two 2D oxides BaO(111) and CoO(220) that have spin splitting. The corresponding band structures calculated with the HSE functional in Fig. 4(g, j) show a clear spin splitting at the K and K' points in the VBM of BaO(111) monolayer and in the vicinity of Y point in the CBM of CoO(220) monolayer. For the former, the splitting reaches around 211 meV, which is much larger than the splitting size ~148 meV of $MoS_2$[40]; for the latter, although the spin splitting is smaller about 21 meV, the large splitting constant ($\alpha = 2E/k = 1.21$ eV·Å) is sufficient to ensure the proper functioning of spintronic devices operating at room temperature.

According to the partial band structures shown in Supplementary Fig. 39, the strong spin splitting in the VBM of BaO(111) is mainly attributed to the O p orbital, while the spin splitting in the CBM of CoO(220) is dominated by the Co d orbital. The spin textures of BaO(111) and CoO(220) obtained from all atoms in the unit cell are shown in Supplementary Figs. 40-41 and exhibit strong spin polarization. For CoO(220) monolayer, the spin-up and spin-down states in the vicinity of Y point in the CBM converge from all directions at the doubly degenerate Y points, exhibiting a Rashba-type spin splitting. In contrast, for BaO(111) monolayer, the spin vectors at the energy valley are perpendicular to the XY plane, which is similar to Zeeman-type splitting occurring in $WSe_2$ induced by the electric field[41].

Finally, we calculate the elastic constants for high-$k$, auxetic, and spintronic 2D oxides obtained above to further assess their resistance ability to the distortions in the presence of strain. The number of independent elastic constants depends on the symmetry of 2D crystals[42]. For example, BaO(111), $HfO_2$(-111) and $ZrO_2$(002)/(220) 2D oxides that belong to hexagonal cell lattice type (in-plane lattice parameters: a=b, γ=120°) have two independent elastic constants: $C_{11}$ and $C_{12}$. These elastic constants satisfy the Born-Huang criteria, $C_{11} > 0$ and $C_{11} > |C_{12}|$, further validating the mechanical stability of BaO(111), $HfO_2$(-111) and $ZrO_2$(002)/(220). The elastic constants and elastic stability conditions for 15 2D oxides are summarized in Supplementary Tables 23-26.

In conclusion, we identify 73 easily/potentially exfoliable 2D oxides from 48 experimentally known non-layered 3D oxides by geometric predicting and vdW-DFT calculations. We develop a generalized geometric algorithm to rapidly filter out the promising exfoliated crystal planes from non-layered precursors belonging to any crystallographic system. The promising planes obtained by geometric consideration are further precisely identified using the binding energies calculated by vdW-DFT. 61 dynamically stable 2D oxides are comprehensively characterized in static dielectric constants, band gaps, spin and mechanical properties, disclosing a variety of functional 2D oxides, including high-$k$ 2D oxides $GeO_2$ (011)/(1-11)/(101) with negative Poisson's ratio, ferromagnetic half semiconductors $VO_2$(022) and $VO_2$(110) monolayers, non-magnetic spintronic materials with significant spin splitting CoO(220) and BaO(111) monolayers, and various auxetic monolayers, such as $WO_3$(110) monolayer, it expands laterally, regardless if a longitudinal stretched or compressed is applied.

## Methods

**Calculation details.** All calculations are performed within density functional theory (DFT)[43] in the Vienna Ab-initio Simulation Package (VASP)[44] using the projector-augmented wave (PAW) pseudopotentials[45] with the generalized gradient approximation GGA/PBE+U[46] exchange-correlation functional. The U values are energy corrections for the spurious self-interaction energy introduced by GGA. Grimme's DFT-D3[24] scheme is adopted to describe the interlayer long-range vdW interactions. Our GGA+U calculations with Grimme's D3 dispersion correction can reasonably reproduce the binding energies for well-known layered materials (such as, graphite, hexagonal boron nitride and several transition-metal dichalcogenides), see Supplementary Fig. 2. The plane-wave energy cutoff of 520 eV is used. During geometry optimization, numerical convergence is achieved with a tolerance of $10^{-5}$ eV in energy and 0.01 eV/Å in force, respectively. To avoid the interaction between a layer and its replica, a vacuum space of 15 Å thickness is added. The 2D Brillouin zone is sampled by a Monkhorst–Pack k-point grid with a uniform spacing of 0.01 Å$^{-1}$ for DOS calculations and 0.02 Å$^{-1}$ for other calculations. The hybrid functional of HSE06[47] is also used to calculate electronic structure in addition to the PBE functional, for improved accuracy. The monolayer structural stability is evaluated by phonon dispersion based on the density functional perturbation theory (DFPT)[48].

**Close-packed degree.** For a truly close-packed plane made up of atoms of diameter D, the area per atom ($A_{CP}$) is given by $A_{CP} = \sqrt{3}/2 \, D^2$. Similarly, we use $A_{hkl}$ to indicate the area per atom for a given crystal plane (h k l). Here, we use the ratio of $A_{CP}$ to $A_{hkl}$ to describe the degree of close-packed for the given crystal plane (h k l). The area per atom in the plane (h k l) is calculated from the following equation[22]:

$$A_{hkl} = \frac{V}{|F_{hkl}| \times d_{hkl}} \tag{1}$$

where V is the volume of unit cell, $|F_{hkl}|$ denotes the structural factor for the plane (h k l), and $d_{hkl}$ is the spacing between adjacent planes.

**Static dielectric tensor.** The static dielectric constant (k) includes both the electronic and the ionic contributions to the dielectric response. We employ density-functional perturbation theory to calculate the static dielectric tensor, from which we extract the in-plane ($\parallel$) and out-of-plane dielectric constants ($\perp$). The in-plane static dielectric constant is obtained by averaging the x and y components, that is, $k_{\parallel} = (k_{\parallel} + k_{\perp})/2$, and the out-of-plane dielectric constant equals to the z component.

Static dielectric tensor of 2D structure computed by DFT contains not only the contribution of material itself but also vacuum layer, due to the fact that the macroscopic electric field is exerted on the supercell containing the vacuum layer. Hence, to obtain material static dielectric constant, we remove the contribution of vacuum layer using the following equations[26]:

$$k_{\parallel}^{2D} = \frac{c}{t}(k_{\parallel}^{sup} - 1) + 1 \tag{2}$$

$$k_{\perp}^{2D} = \left[\frac{c}{t}\left(\frac{1}{k_{\perp}^{sup}} - 1\right) + 1\right]^{-1} \tag{3}$$

where $c$ is the supercell height, and $t$ is the thickness of monolayer. The thickness $t$ is estimated by the interlayer distance of the bilayer. Besides, we also calculate static dielectric constants of monolayers with vacuum size of 30 Å, which are in good agreement with that of 15 Å, see Supplementary Fig. 32.

**Figure-of-merit.** The gate leakage current density ($J_G$) caused by direct-tunneling can be expressed by the following semi-empirical formula[48]:

$$J_G \propto \exp\{-\frac{4\pi\sqrt{2|q|}\cdot(m_{eff}\Phi_b)^{1/2}\varepsilon_0 \cdot t_{ox,eq}}{h}\} \tag{4}$$

where $q$ and $m_{eff}$ are the charge and tunneling effective mass respectively for the electron or hole; $\Phi_b$ is the injection barrier; $\varepsilon_0$ is static dielectric constant, $t_{ox,eq}$ represents the equivalent oxide thickness, and $h$ is the Planck's constant.

Formula (4) indicates that the leakage current at a given thickness decreases exponentially with the increase of the dielectric parameters ($m_{eff}$, $\Phi_b$ and $\varepsilon_0$). For convenience, $(m_{eff}\Phi_b)^{1/2}\varepsilon_0$ in the exponent is defined as figure-of-merit $f_0$. Of course, one can compute $m_{eff}$ and $\Phi_b$ accurately, but considering of the computational burden, here we do a rough but reasonable simplification for figure-of-merit, i.e., the product of $m_{eff}$ and $\Phi_b$ is roughly proportional to $E_g$, which approximates figure-of-merit as simply $E_g \cdot \varepsilon_0$, similar approach can be found in the literatures[49,50].